\documentclass[aps,prl,twocolumn,superscriptaddress,showpacs,floatfix,longbibliography]{revtex4-2}
\pdfoutput=1
\usepackage{amsmath,amssymb,amsfonts,float,graphics,epsfig,epstopdf,color,verbatim,tabularx,bm,multirow,appendix,hyperref}
\usepackage[normalem]{ulem}
\usepackage{lmodern}
\usepackage{ytableau}
\usepackage{pifont}
\usepackage{color}
\usepackage{bm}
\usepackage{wasysym}

\usepackage{ dsfont }
\usepackage{graphicx}

\def\k{{\mathbf{k}}}
\def\a{{\mathbf{a}}}

\def\G{{\mathbf{G}}}
\def\K{{\mathbf{K}}}

\def\q{{\mathbf{q}}}

\def\bK{{\boldsymbol{\kappa}}}

\def\r{{\mathbf{r}}}

\def\C{{\mathcal{C}}}
\def\P{{\mathcal{P}}}
\def\T{{\mathcal{T}}}
\def\moire{{moir\'e }}
\def\hk{{\hat{\mathbf{k}}}}

\newcommand{\<}{\langle}

\renewcommand{\>}{\rangle}

\renewcommand{\b}[1]{\mathbf{#1}} 

\begin{document}

\title{Multi-Q spin-valley order in twisted WSe$_2$}

\author{Arthur Bril}
\affiliation{Department of Physics and Astronomy, Ghent University, Krijgslaan 281, 9000 Gent, Belgium}

\author{Nai Chao Hu}
\affiliation{Department of Physics and Astronomy, Ghent University, Krijgslaan 281, 9000 Gent, Belgium}

\author{Nick Bultinck}
\affiliation{Department of Physics and Astronomy, Ghent University, Krijgslaan 281, 9000 Gent, Belgium}

\begin{abstract}
We report on a study of the interacting phase diagram of $3.65^\circ$-twisted WSe$_2$ at moir\'e hole filling $\nu=1$, in which we find previously-overlooked types of magnetism. Specifically, in part of the phase diagram we obtain a magnetic order parameter which modulates in space with four different non-zero wave vectors, corresponding to the three $M$-points and one $K$-point of the \moire Brillouin zone. These multi-Q orders, which can be coplanar or non-coplanar, are continuous deformations of the $120^\circ$ spin-valley anti-ferromagnet (AFM), where the unit cell has expanded by a factor of four. Interestingly, we find that the multi-Q states are stabilized for experimentally relevant values of interaction strength and displacement field, and are accompanied by a softening of the spin fluctuations near the $M$-points of the \moire Brillouin zone.
\end{abstract}
\date{\today}
\maketitle

\emph{Introduction --} Moir\'e superlattice systems have opened up a new pathway for the study of strongly-correlated electron systems in two dimensions. An interesting subclass of these \moire materials consists of twisted transition metal dichalcogenides (TMD's). Twisted TMD's combine strong spin-orbit coupling with non-trivial band topology, and allow for a high tunability of \moire band structures by applying an out-of-plane displacement field, making them an ideal platform for studying strongly-correlated quantum matter.

In this work we numerically study the interacting phase diagram of twisted bilayer WSe$_2$ (tWSe$_2$) at twist angle $3.65^{\circ}$ and \moire hole filling $\nu = 1$, where recent experiments have found superconductivity at small displacement fields~\cite{xia2025superconductivity}. Superconductivity has subsequently also been observed in larger twist-angle devices at hole fillings slightly shifted away from $\nu=1$, and at higher displacement fields~\cite{Guo_2025,xia2025hightcmoire}. The phase diagram of tWSe$_2$ in the vicinity of the superconducting dome shares many similarities with that of high-temperature cuprate superconductors~\cite{xia2025hightcmoire}, which has spurred a variety of theoretical works ~\cite{belanger2022superconductivity, klebl2023competition, wu2023pair, zegrodnik2023mixed, schrade2024nematic, akbar2024topological, chubukov2025quantum, Kim_2025, Zhu_2025, 7z4z-vlj8, Xie_2025, guerci2024topological, tuo2024theory, qin2025topological, fischer2024theory, peng2025magnetism, xie2025kondo, ryee2025site,Munoz2025}.

Below we uncover a new type of magnetic order in the zero-temperature phase diagram of tWSe$_2$, characterized by intricate spin textures that modulate in space with four different wave vectors. These multi-Q magnetic orders arise in a region of the phase diagram close to where superconductivity is observed, and form out of the $120^\circ$ anti-ferromagnet (AFM) via a continuous phase transition. The approach to this phase transition is accompanied by a gradual softening of the spin fluctuations near the $M$-points of the \moire Brillouin zone, which we expect to play an important role for superconductivity. By varying the interaction strength and displacement field we find two distinct multi-Q orders, with one being coplanar and the other one non-coplanar. The system remains incompressible at the transition between the $120^\circ$ AFM and the non-coplanar order, implying that the latter has zero Chern number. We check the robustness of our results by using two different tWSe$_2$ continuum model, which we now introduce.

\emph{Continuum models --} The generic continuum model for a twisted homo-bilayer TMD in valley $+K$ reads
\begin{align}
    H^{+K}(\b{r}) = \begin{pmatrix}
    T_1 + \Delta_1(\b{r}) & \Delta_T(\b{r}) \\
    \Delta_T^{\dagger}(\b{r}) & T_2 + \Delta_2(\b{r})
    \end{pmatrix}\,,\label{eq:cm1}
\end{align}
where the matrix structure is in layer space. The Hamiltonian for the $-K$ valley is related by time-reversal symmetry, $H^{-K}(\b{r}) = \T H^{+K}(\b{r}) \T^{-1}$. Due to the large Ising spin-orbit coupling there is spin-valley locking \cite{PhysRevB.84.153402}, and hence we will use spin and valley interchangeably in this work.  Single-particle intervalley couplings are neglected due to the large separation between the \moire and atomic length scales~\cite{fengcheng2019homo}.

The kinetic energy $T_l$, the potentials $\Delta_l(\r)$ and the inter-layer tunneling amplitude $\Delta_T(\r)$ are constrained by the symmetries of the system, which include a three-fold rotation $\C_{3z}$, and the $\C_{2y}\T$ symmetry. Importantly, the anti-unitary symmetry $\C_{2y}\T$ does not preclude a non-zero (valley) Chern number, as the action of the $\C_{2y}$ is improper, so that the valley-resolved Chern number is even under $\C_{2y}\T$. Indeed, the quantum anomalous Hall responses have already been explored when the model in~\eqref{eq:cm1} was first introduced in Ref.~\cite{fengcheng2019homo}, and have now been observed in different experimental setups \cite{cai2023signatures, foutty2024mapping, zeng2023thermodynamic, park2023observation, xu2023observation, kang2024double, kang2024evidence}. 

To ensure that our results hold with sufficient generality, we use two different continuum models. The first continuum model uses a lowest-order approximation for $T_l$, $\Delta_l(\r)$ and $\Delta_T(\r)$, as in~\cite{fengcheng2019homo}. This Lowest-Order (LO) model contains only three parameters, for which we use the values of Ref.~\cite{devakul2021magic}. One caveat in using these parameters, however, is that they were extracted from DFT simulations on a $5^\circ$ twist-angle structure, whereas here we aim to study a smaller twist angle $\theta = 3.65^\circ$.

The second continuum model we use is based on Ref.~\cite{zhang2024universal}, where the authors took both higher \moire-harmonics and higher-order gradient terms into account, and parameters were fitted to relaxed DFT band structures obtained at $\theta = 3.48^\circ$. The original Higher-Order (HO) continuum model of Ref.~\cite{zhang2024universal} contains $77$ parameters. In this work, however, we use an approximation which keeps only $16$ parameters, corresponding to the first \moire harmonics (but including higher-order gradient terms). The detailed expressions for $T_l$, $\Delta_l(\r)$ and $\Delta_T(\r)$ in the HO model can be found in the supplementary material~\cite{supp}. There we also show that the $16$-parameter model is able to reproduce the bands from the original $77$-parameter model to very high accuracy.

\emph{Interacting phase diagram --} We make the model interacting by adding a density-density interaction with a gate-screened Coulomb potential. Concretely, in momentum space the interaction potential is written as $V(\q) = \frac{e^2}{2\epsilon_0\epsilon}\tanh(qD)/q$, where $D$ is the sample-gate distance, and $\epsilon$ is the relative dielectric constant determined by the dielectric environment of the sample. We fix $D = 50 a_{0} \approx 16.5$nm, with $a_{0}$ the WSe$_{2}$ lattice constant, and take $\epsilon$ to be a tuning parameter. As the WSe$_2$ continuum model is obtained by fitting to DFT bands, it already incorporates some effects of the interaction. Therefore, to reduce the double counting of interaction effects, we add quadratic `subtraction terms' to the Hamiltonian which ensure that the continuum model bands constitute an exact solution of the Hartree-Fock (HF) self-consistency equations. For our numerical simulations we keep 4 active bands per valley. The remaining bands are assumed to be completely occupied. A detailed derivation of the complete Hamiltonian is given in the supplementary material~\cite{supp}.

In experiment, the tWSe$_2$ sample is encapsulated in a dual-gated device, which can be used to apply an electric displacement field $E$ perpendicular to the WSe$_2$ layers. We incorporate the displacement field in the continuum model by adding a potential energy difference between the top and bottom layers given by $u_{D} = Ed\frac{\epsilon_{\text{hBN}}}{\epsilon_{\text{TMD}}}$, where $d \approx 0.7$ nm is the distance between the layers, $\epsilon_{\text{hBN}} \approx 3$ and $\epsilon_{\text{TMD}} \approx 8$ \cite{xia2025superconductivity}. Our main focus is the phase diagram at hole doping $\nu=1$, i.e. one hole per \moire unit cell, as a function of both $E$ and $\epsilon$. In the main text we show the results for the HO model, and comment on the quantitative differences that occur in the LO model. The detailed numerical results for the LO model can be found in the supplementary material.

\begin{figure*}[t!]
    \centering
    \includegraphics[width=\linewidth]{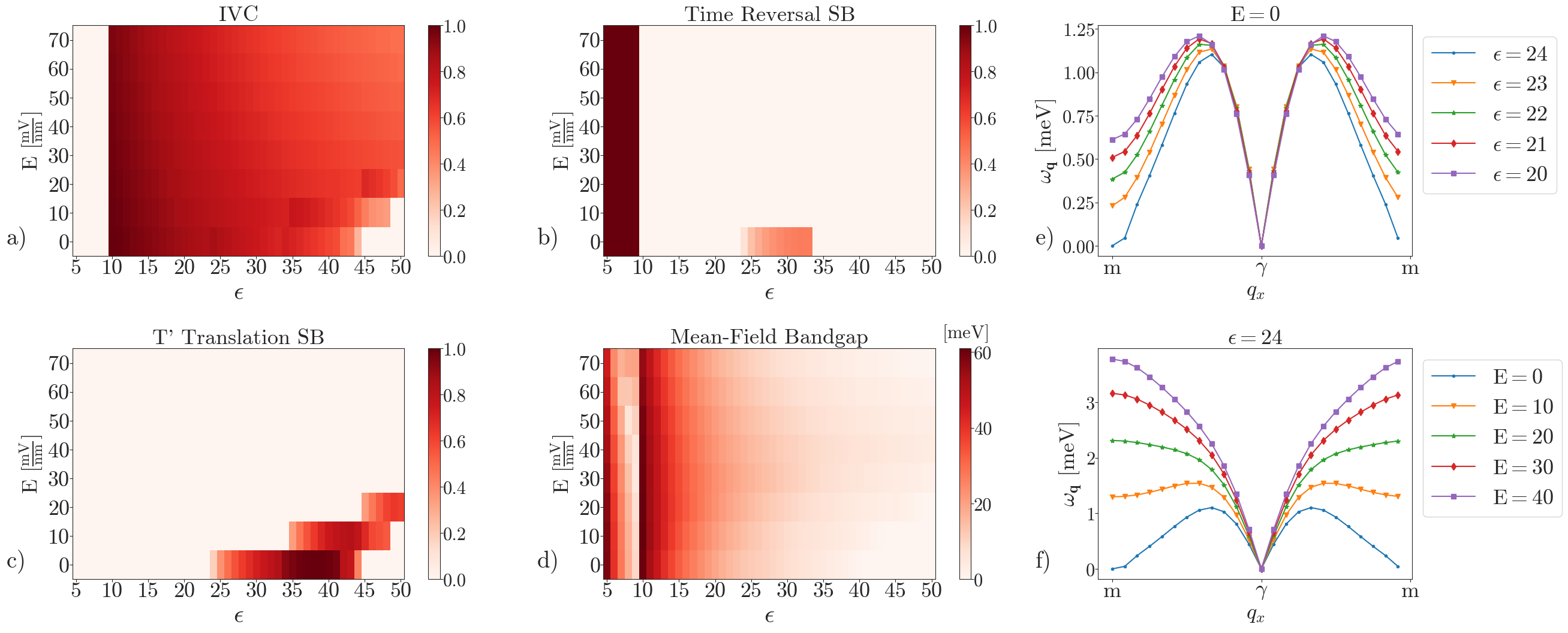}
    \caption{Numerical results for HO model obtained on a $24\times 24$ \moire lattice. (a) Spin-valley U(1) order parameter $O_{\text{IVC}}$, (b) $\mathcal{T}'$ time-reversal order parameter $O_{\mathcal{T}'}$, (c) order parameter $O_{T'}$ for the generalized translation symmetry $T'_{\a_i}$. (d) Mean-field bandgap. (e-f) TDHF Goldstone mode dispersion relation $\omega_\q$ along $\q=(q_x,q_y=0)$, both for fixed $E$ as a function of $\epsilon$ (e), and for fixed $\epsilon$ as a function of $E$ (f).}
    \label{fig:BigFig_HO}
\end{figure*}

From previous works~\cite{Wang2023,fischer2024theory,tuo2024theory,peng2025magnetism,Munoz2025}, we know that part of the phase diagram hosts a $120^\circ$ anti-ferromagnet (AFM). On the \moire scale, the AFM modulates with a wave vector $\boldsymbol{\kappa}_+$ or $\boldsymbol{\kappa}_-$, with $\boldsymbol{\kappa}_\pm$ the $K$-points of the \moire- or mini-Brillouin zone (mBZ) (see~\cite{supp} for definition). Exactly which $K$-point determines the spin modulation depends on the orientation of the displacement field (at $E=0$ the wave vector is determined via a spontaneous breaking of the $\mathcal{C}_{2y}$ symmetry). The $120^\circ$ AFM is invariant under $T'_{\a_i} = e^{-\frac{i}{2}\boldsymbol{\kappa}_\pm \cdot \a_i\sigma^z} T_{\a_i}$, where $\a_i$ are \moire lattice vectors, and $T_{\a_i}$ is the translation operator: $T^{-1}_{\a_i}\psi^\dagger(\r)T_{\a_i} = \psi^\dagger(\r+\a_i)$. So $T'_{\a_i}$ is a combination of a translation and a spin rotation along the $z$-axis. To exploit this symmetry in our numerical simulations we work in the Bloch basis associated with $T'_{\a_i}$ (i.e. the single-particle states are labeled with $\tilde{\k}$, which refers to the $T'_{\a_i}$ eigenvalues $e^{i\tilde{\k}\cdot \a_i}$). 

The $120^\circ$ AFM is also invariant under $\mathcal{T}' = \sigma^z \mathcal{T} = \sigma^x \mathcal{K}$, where $\mathcal{K}:i\rightarrow -i$ represents complex conjugation. This time-reversal symmetry is a combination of the standard Kramers time-reversal symmetry $\mathcal{T}=i\sigma^y \mathcal{K}$ and a $\pi$ spin rotation along the $z$-axis. The $\mathcal{T}'$ symmetry leaves the in-plane XY components of the spin invariant, but flips the out-of-plane Z component. The AFM is thus compatible with the $\mathcal{T}'$ symmetry because it is coplanar.

From the numerical solution of the HF self-consistency equations we obtain
\begin{equation}
\left[P(\tilde{\k},\tilde{\k}')\right]_{\alpha\sigma;\alpha'\sigma'} := \langle \psi^\dagger_{\alpha'\sigma'}(\tilde{\k}')\psi_{\alpha\sigma}(\tilde{\k}) \rangle\,,
\end{equation}
where $\psi^\dagger_{\alpha\sigma}(\tilde{\k})$ creates an electron with spin $\sigma$ in band $\alpha$ of the WSe$_2$ continuum model. From $P$, we define an order parameter for Inter-Valley Coherence (IVC), i.e. U(1) spin or valley symmetry breaking, as $O_{\text{IVC}} := N_{\text{UC}}^{-1}\sum_{\tilde{\k},\tilde{\k'}} \sum_{\alpha,\alpha'} \big|\left[P(\tilde{\k},\tilde{\k}') \right]_{\alpha\uparrow;\alpha'\downarrow}\big|^2 $, with $N_{\text{UC}}$ the number of \moire unit cells. Similarly, we define an order parameter $O_{T'} := N_{\text{UC}}^{-1}\sum_{\tilde{\k},\tilde{\k'}} (1-\delta_{\tilde{\k},\tilde{\k}'})|| P(\tilde{\k},\tilde{\k}')||^2 $ for $T'_{\a_i}$ translation-symmetry breaking, and $O_{\mathcal{T}'} := N_{\text{UC}}^{-1}\sum_{\tilde{\k},\tilde{\k'}} ||P(\tilde{\k},\tilde{\k}') - \sigma^x P^*(-\tilde{\k},-\tilde{\k}')\sigma^x||^2$ for $\mathcal{T}'$ time-reversal symmetry breaking ($||\cdot||$ is the Frobenius norm).

Figs.~\ref{fig:BigFig_HO} (a-d) show the interacting phase diagram in terms of these order parameters, as well as the mean-field bandgap and Goldstone mode dispersion. All results were obtained from numerical solutions of the HF self-consistency equations for a system with $24\times 24$ \moire unit cells.

For small $\epsilon$ ($\epsilon\lesssim 24)$, we obtain the spin-Valley Polarized (VP) state and the $120^\circ$ AFM reported in previous works~\cite{Wang2023,fischer2024theory,tuo2024theory,peng2025magnetism,Munoz2025}. The VP state is diagnosed by $O_{\mathcal{T}'}\neq 0$ and $O_{\text{IVC}}=O_{T'}=0$, whereas the $120^\circ$ AFM has $O_{\text{IVC}} \neq 0$ and $O_{\mathcal{T}'}=O_{T'} = 0$. The VP and AFM states are separated by a first-order transition at $\epsilon\sim 9$.

For larger values of $\epsilon$ we find IVC states, i.e. states with non-zero $O_{\text{IVC}}$, which differ from the $120^\circ$ AFM because they break the generalized translation symmetry $T'_{\a_i}$ ($O_{T'}\neq 0$). These IVC states are the multi-Q orders which, as we show below, host spin textures that indeed modulate in space with four different wave vectors. In Fig.~\ref{fig:BigFig_HO} (d) we see that in most of the phase diagram the multi-Q orders are incompressible, albeit with a smaller charge gap than the $120^\circ$ AFM. When the system becomes metallic at the largest values of $\epsilon$ used in this work (e.g. $\epsilon=41$ at $E=0$), the IVC and $T'_{\a_i}$ order parameters quickly drop to zero, after which a symmetric metal is formed.

The transition from the $120^\circ$ AFM to the multi-Q order is continuous, irrespective of whether the transition is driven by changing $\epsilon$ or $E$. A first indication of this can be seen by looking at the mean-field bandgap, which does not close and evolves smoothly across the transition. In Figs.~\ref{fig:BigFig_HO} (e-f) we also show the energy $\omega_\q$ of the $120^\circ$ AFM Goldstone mode along $(q_x,q_y=0)$, obtained via the RPA Bethe-Salpeter equation (or equivalently via the Time-Dependent HF (TDHF) equation), as reviewed in Ref.~\cite{Khalaf2020}. Fig.~\ref{fig:BigFig_HO} (e) shows $\omega_\q$ at $E=0$ as a function of $\epsilon$. When the transition to the multi-Q order at $\epsilon\sim 24$ is approached, the Goldstone mode at the $M$-points of the mBZ continuously lowers in energy, until the gap closes at the transition (the plot shows only one $M$-point, but $\mathcal{C}_{3z}$ symmetry is preserved, and identical behaviour is therefore found at all three $M$-points). In Fig.~\ref{fig:BigFig_HO} (f) a similar softening of the Goldstone mode energy near the $M$-points can be seen to occur as a function of $E$, working at fixed $\epsilon=24$. As suggested by this condensation of the Goldstone mode, we indeed find that the multi-Q orders are characterized by four different wave vectors: the $\boldsymbol{\kappa}_\pm$ wave vector of the original $120^\circ$ state, and the three $M$-point wave vectors.

From Figs.~\ref{fig:BigFig_HO} (b-c) we see that there are two distinct multi-Q magnetic orders: there is a $\mathcal{T}'$-breaking variant with non-coplanar spin order, and a $\mathcal{T}'$-symmetric variant with coplanar order. The non-coplanar multi-Q order appears at smaller values of $\epsilon$. The transition from non-coplanar to coplanar order occurs at $\epsilon = 33$ in the absence of a displacement field. For non-zero $E$, the transition shifts to higher $\epsilon$. 

\begin{figure*}[t!]
    \centering
    \includegraphics[width=\linewidth]{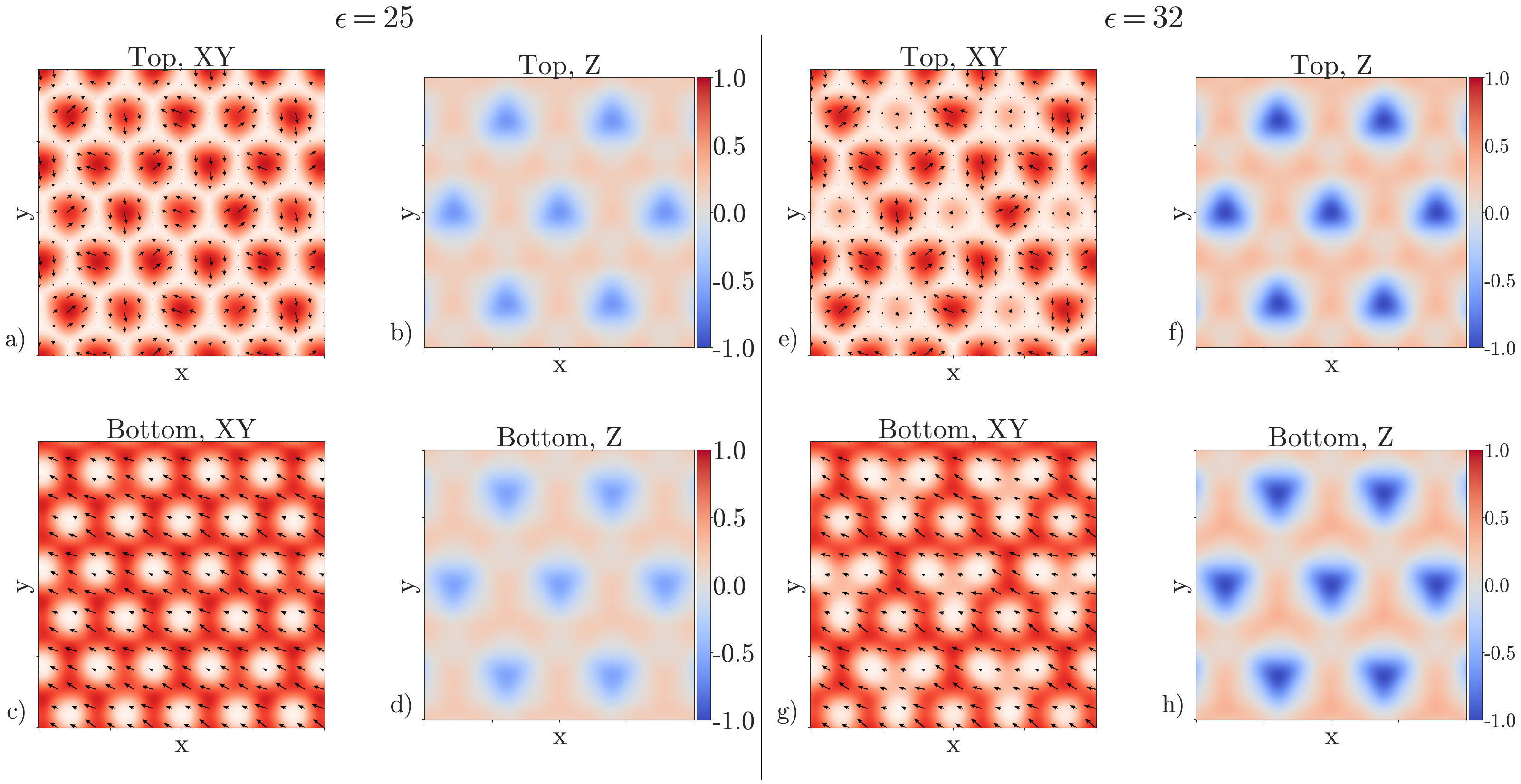}
    \caption{(a-d) Non-coplanar multi-Q spin texture in the top and bottom layers at $\epsilon = 25$. (a) In-plane magnetic order in the top layer, described by $|M_{xy}^t(\r)|$ (color) and $\theta_t(\r)$ (arrows). (b) $M_z^t(\r)$, the out-of-plane component of magnetic order in the top layer. (c) In-plane magnetic order in the bottom layer, described by $|M_{xy}^b(\r))|$ (color) and $\theta_b(\r)$ (arrows). (d) $M_z^b(\r)$, the out-of-plane component of magnetic order in bottom layer. (e-h) Same as (a-d), but for $\epsilon=32$.}
    \label{fig:spintext_HO}
\end{figure*}

For the HO model used in the main text, the non-coplanar order is only found in a relatively small region of the phase diagram, corresponding to $24\lesssim \epsilon \lesssim 33$ and $E\lesssim 10 $ mV/nm. In contrast, for the LO model discussed in the Supplementary Material, the non-coplanar order is the dominant multi-Q state, and the coplanar order is only found in a small region of the phase diagram at high $\epsilon$. But otherwise both phase diagrams are very similar: they host the same broken-symmetry orders, and nearly identical multi-Q states are found in both the HO and LO model.

Spin textures for the non-coplanar multi-Q state at two representative values of $\epsilon$ are shown in Fig.~\ref{fig:spintext_HO}. We plot $|M^l_{xy}(\r)|$, $\theta_l(\r)$ and $M^l_z(\r)$, which are defined as
\begin{eqnarray}
\langle \psi^\dagger_{l,\uparrow}(\r)\psi_{l,\downarrow}(\r)\rangle & = & M^l_x(\r) + iM^l_y(\r) \nonumber \\
 & =: & e^{-2i\K_l\cdot\r}|M^l_{xy}(\r)| e^{i\theta_l(\r)} \\
\langle \psi^\dagger_l(\r)\sigma^z \psi_l(\r)\rangle & = & M_z^l(\r)
\end{eqnarray}
Here, $l=t/b$ is the layer index, and $\K_{t/b} = R(\pm \theta/2)\K$ are the $K$-points of the mono-layer Brillouin zones in the top and bottom layers (which are rotated over respectively $+\theta/2$ and $-\theta/2$). The colors in Fig.~\ref{fig:spintext_HO} represent $|M_{xy}^l(\r)|$ and $M_z^l(\r)$. The in-plane direction of the magnetic order, i.e. $\theta_l(\r)$, is represented by black arrows.

For $\epsilon = 25$ (Figs.~\ref{fig:spintext_HO} (a-d)), which is close to the onset of the non-coplanar order, the in-plane spin order for the top layer closely resembles that of the $120^{\circ}$ state, while for the bottom layer a nearly constant $\theta_b(\r)$ is found. For $\epsilon=32$ (Figs.~\ref{fig:spintext_HO} (e-h)), we see a clear doubling of the periodicity along all \moire lattice directions, leading to a unit cell that is four times larger than the unit cell of the $120^\circ$ state. Fig.~\ref{fig:spintext_HO} (e) shows that the $120^\circ$ in-plane order on the top layer becomes significantly weaker in $1/4$ of the \moire unit cells, but survives on a Kagome lattice made out of the remaining $3/4$ of \moire unit cells. Concomitantly, out-of-plane order ($M_z^l(\r)\neq 0)$ develops on the $2\times 2$ triangular lattice consisting of the \moire unit cells with negligible in-plane order, as shown in Figs.~\ref{fig:spintext_HO} (f) and (h) for both top and bottom layer.

\begin{figure}
    \centering
    \includegraphics[width=\linewidth]{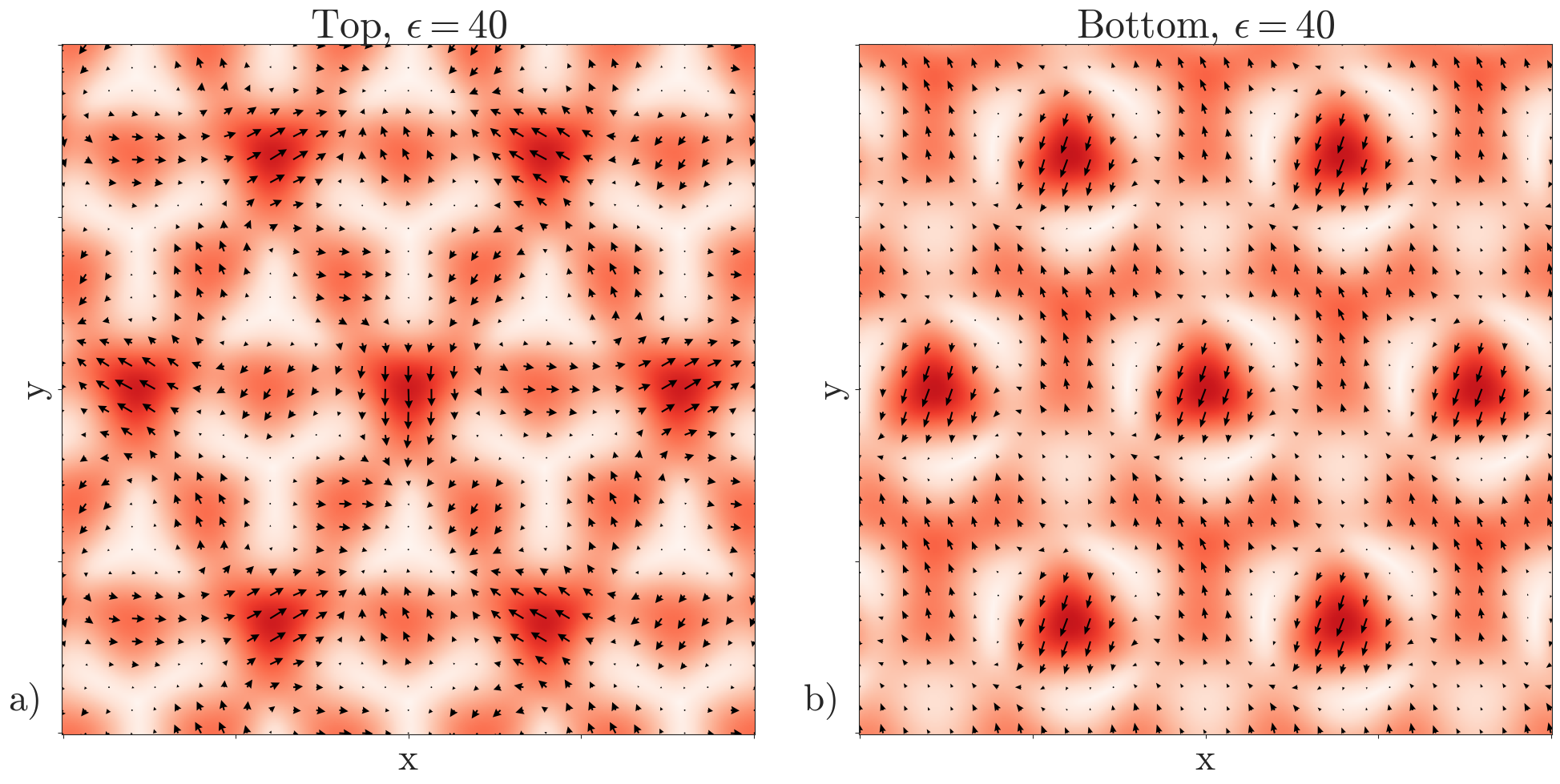}
    \caption{In-plane spin texture for the coplanar multi-Q state at $\epsilon=40$ in (a) top layer and (b) bottom layer. Color represents $|M_{xy}^l(\r)|$, and arrows represent $\theta_l(\r)$.}
    \label{fig:HO_coplanar}
\end{figure}

The spin texture for the coplanar multi-Q state at $\epsilon=40$ is shown in Fig.~\ref{fig:HO_coplanar}. The in-plane spin order again forms a $120^{\circ}$ state, but on a triangular lattice with a periodicity of \emph{two} \moire lattice constants. More surprising is that the spin order on the bottom layer, as shown in Fig.~\ref{fig:HO_coplanar} (b), is no longer uniform, but consists of isolated patches which have a different spin orientation than their environment.

\emph{Conclusions --} We have shown that at moderately large values of the relative dielectric constant, and for small displacement fields, the $120^\circ$ AFM in tWSe$_2$ at $\nu = 1$ hole doping becomes unstable towards the formation of a multi-Q magnetic order with enlarged spatial unit cell. Ref.~\cite{peng2025magnetism} has estimated the appropriate value for the dielectric constant to be $\epsilon \sim 30$, by comparing the layer-polarized regime in Hartree-Fock with the layer-polarized regime observed in experiment. For this value of $\epsilon$ we find a multi-Q spin order in the HO model. For both the HO and LO model, $\epsilon=30$ is close to the critical value at which the $120^\circ$ state becomes unstable to the multi-Q order. 

Our TDHF simulations show that the transition from the $120^\circ$ state to the multi-Q order is continuous, and happens via a gradual softening of the spin fluctuations at the $M$-points of the mini-Brillouin zone. A recent experiment~\cite{Xiong2025} has observed two types of neutral collective modes in tWSe$_2$: one fast mode, corresponding to the spin or valley Goldstone mode, and one slow mode. The slow mode was interpreted as a gapped amplitude mode~\cite{Xiong2025}. Our results provide an alternative interpretation, however: the slow mode likely corresponds to the soft spin fluctuations near the $M$-points. In our simulations, the gap of the $M$-point spin fluctuations is less than $1$ meV, and thus an order of magnitude smaller than the gap of the amplitude mode, which is comparable to the mean-field bandgap $\sim 5-10$ meV.

Interestingly, the multi-Q orders appear around zero displacement field, which is where superconductivity is observed in experiment \cite{xia2025superconductivity,xia2025hightcmoire}. So even if no static multi-Q order develops in the tWSe$_2$ samples, the softening of the spin fluctuations near the $M$-points could still play an important role for Cooper pairing. One interesting scenario is that the multi-Q order is preempted by the formation of superconductivity, which is stabilized by the soft spin fluctuations near the $M$-points. We will investigate this scenario in future work.

\begin{acknowledgements}
{\noindent \it Acknowledgements.---} We thank Taige Wang for helpful discussions and Liangtao Peng for additional explanation of the results in~\cite{peng2025magnetism}. This research was supported by the European Research Council under the European Union Horizon 2020 Research and Innovation Programme via Grant Agreement No. 101076597-SIESS (N.H. and N.B.), and by a grant from the Simons Foundation (SFI-MPS-NFS-00006741-04) (A.B and N.B.).
\end{acknowledgements}

\bibliographystyle{apsrev}
\bibliography{bib}

\clearpage
\onecolumngrid

\begin{center}
    \textbf{\large Supplementary Material} 
    
    \vspace{0.5 cm}
    
   \textbf{\large Multi-Q spin-valley order in twisted WSe$_2$}
    
    \vspace{0.5 cm}
    
    Arthur Bril$^1$, Nai Chao Hu$^1$, Nick Bultinck$^1$
    
    \vspace{0.2 cm}
    
    \textit{$^1$Department of Physics and Astronomy, Ghent University, Krijgslaan 281, 9000 Ghent, Belgium}
    
    \vspace{0.5 cm}

    (Dated: \today)
\end{center}

\section{Continuum model band structures}\label{app:bs}
As explained in the main text, the general $K$-valley continuum Hamiltonian for twisted TMD's takes on the form
\begin{align}
    H^{+K}(\b{r}) = \begin{pmatrix}
    T_1 + \Delta_1(\b{r}) & \Delta_T(\b{r}) \\
    \Delta_T^{\dagger}(\b{r}) & T_2+ \Delta_2(\b{r})
    \end{pmatrix}\label{eq:cm}
\end{align}
The matrix structure is in layer space. Below we give the detailed expression for $T_l$, $\Delta_l(\r)$ and $\Delta_T(\r)$ in the Lowest-Order (LO) and Higher-Order (HO) continuum model.

\subsection{Lowest-order model}
In the LO model, we keep terms to leading order in $\k$: The kinetic energy in layer $l$ is described by a quadratic dispersion, $T_l = -\hbar^2(\hat{\k}-\bK_l)^2/2m^*$, centered at the \moire Brillouin zone corner $\bK_l$. The intra-layer potentials are written as
\begin{equation}
\Delta_l(\hat{\b{r}}) = 2V\sum_{j=1,3,5}\cos(\b{G}_j\cdot\hat{\b{r}}+(-)^l\phi)\,,
\end{equation}
and the inter-layer tunneling as
\begin{equation}
\Delta_T(\hat{\b{r}}) = w(1+e^{-i\b{G}_2\cdot\hat{\b{r}}}+e^{-i\b{G}_3\cdot\hat{\b{r}}})\,,
\end{equation}
as in Ref.~\cite{fengcheng2019homo}, where $\b{G}_j$ are the \moire reciprocal lattice vectors, defined as $\frac{(j-1)\pi}{3}$ counter clockwise rotations of $\b{G}_{1} = \left(\frac{4\pi\theta}{\sqrt{3}a_{0}},0 \right)$ with $\theta$ the twist angle in radians, and $a_{0}$ the WSe$_2$ lattice constant. The K-points of the mini Brillouin zone are at $\boldsymbol{\kappa}_{+} = \left( -\b{G}_{1} + \b{G}_{3}\right)/3$ and $\boldsymbol{\kappa}_{-} = \left( -\b{G}_{1} - \b{G}_{2}\right)/3$. We adopt the commonly-used set of parameters $(V, \phi, w) = (9 \text{meV}, 128^{\circ}, 18 \text{meV})$ \cite{devakul2021magic}.
\\ \\
It is worth mentioning that this lowest-order model has an accidental \emph{intra-valley} inversion symmetry $\tilde{\P}$, which consists of $\b{r} \mapsto -\b{r}$ followed by a layer exchange. This symmetry, together with $\T$, ensures a two-fold valley degeneracy at each $\k$. This accidental symmetry, combined with $\C_{3z}$, leads to an effective six-fold rotational symmetry in the properties of the electronic wavefunctions. Note that $\tilde{\P}$ is different from a physical global inversion symmetry, which would require $\Delta_T(\b{r}) = 0$, as is the case for the AB-stacked TMD homobilayers.

\subsection{Higher-order model}
\begin{figure}
    \centering
    \includegraphics[width=0.5\linewidth]{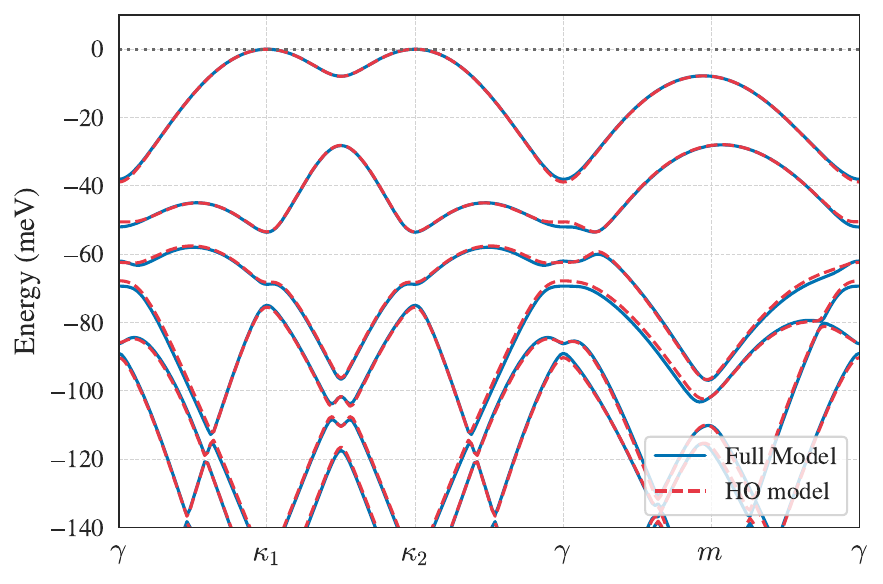}
    \caption{Band structure comparison at $\theta = 3.65^{\circ}$.}
    \label{fig:365}
\end{figure}

In the reduced HO model, simplified from the ``full'' model in Ref.~\cite{zhang2024universal}, the entries of the continuum Hamiltonian are written as
\begin{align}
    T_l =& \frac{\hbar^2}{2m^{11}_l} (\hk-\bK_l)_z(\hk-\bK_l)_{z^*} + \frac{\hbar^2}{2m^{30}_l} \left[(\hk-\bK_l)_z^3+(\hk-\bK_l)_{z^*}^3\right] + \frac{\hbar^2}{2m^{22}_l} (\hk-\bK_l)_z^2(\hk-\bK_l)_{z^*}^2 \nonumber\\
    &+ \frac{\hbar^2}{2m^{41}_l} \left[(\hk-\bK_l)_z^4(\hk-\bK_l)_{z^*}+(\hk-\bK_l)_z(\hk-\bK_l)_{z^*}^4\right] + \frac{\hbar^2}{2m^{33}_l} (\hk-\bK_l)_z^3(\hk-\bK_l)_{z^*}^3 \nonumber\\
    &+ \frac{\hbar^2}{2m^{60}_l} \left[(\hk-\bK_l)_z^6+(\hk-\bK_l)_{z^*}^6\right], \label{eq:tl}\\
    \Delta_l =& \sum_{j=1}^6 e^{i\b{G}_j\cdot\hat{\b{r}}}\big[V_{l,\b{G}_j}^{00} + V_{l,\b{G}_j}^{01}(\hk-\bK_l)_{z^*} + V_{l,\b{G}_j}^{10}(\hk-\bK_l)_{z} + V_{l,\b{G}_j}^{02}(\hk-\bK_l)_{z^*}^2 \nonumber\\
    &\qquad \qquad \quad + V_{l,\b{G}_j}^{11}(\hk-\bK_l)_z(\hk-\bK_l)_{z^*} + V_{l,\b{G}_j}^{20}(\hk-\bK_l)_{z}^2\big],\label{eq:dl}\\
    \Delta_T =& \sum_{j=1}^3 e^{i(\b{q}_1-\b{q}_j)\cdot\hat{\b{r}}}\left[w_{12,\b{q}_j}^{00} + w_{12,\b{q}_j}^{10}(\hk-\bK_2)_{z} + w_{12,\b{q}_j}^{20}(\hk-\bK_2)_{z}^2 + w_{12,\b{q}_j}^{11}(\hk-\bK_2)_z(\hk-\bK_2)_{z^*}\right], \label{eq:dt}
\end{align}
where $(\hk-\bK_l)_z = (\hk-\bK_l)_x + i (\hk-\bK_l)_y$, $(\hk-\bK_l)_{z*} = (\hk-\bK_l)_x - i (\hk-\bK_l)_y$, and $\b{q}_1 = \bK_2 - \bK_1$. 
We can write a term of order $(m,n)$ in $\Delta_{l/T}$ generically as $e^{i\b{q}\cdot\hat{\b{r}}}p_{l'l,\b{q}}^{mn}f_{l'l}^{mn}(\hat{\k})$, where $\b{q}$ is the momentum translation involved, $p_{l'l,\b{q}}^{mn}$ the continuum model parameter, and $f^{mn}_{l'l}(\hat{\k}) = (\hk-\bK_l)^m_z(\hk-\bK_l)^n_{z*}$. In a plane wave basis,
\begin{align}
    \<\k + \b{G}', l'| e^{i\b{q}\cdot\hat{\b{r}}}p_{l'l,\b{q}}^{mn}f_{l'l}^{mn}(\hat{\k}) |\k + \b{G}, l\> = \delta_{\b{G}'-\b{q}, \b{G}}p_{l'l,\b{q}}^{mn}f_{l'l}^{mn}(\k + \b{G}).
\end{align} 
Symmetries $\C_{3z}, \C_{2y}\T$, and hermiticity can relate terms of the expansion within the same order of $\k$ by changing $p_{l'l,\b{q}}^{mn}$ and $f^{mn}_{l'l}$. In practice, the \moire induced part of the Hamiltonian $\Delta = H^{+K}-\text{diag}\{T_1, T_2\}$ can be generated with only the $j=1$ components via 
\begin{equation}
\Delta = \sum_{m=0}^1(\C_{2y}\T)^m\left[\sum_{n=0}^2\C_{3z}^n\left(\Delta(j=1)\right)\right]+h.c.\,, 
\end{equation}
which is then explicitly symmetric and hermitian. 
\\ \\
Compared to the LO model, the higher-order gradient terms spoil the accidental intra-valley inversion symmetry $\tilde{\P}$, resulting in a valley splitting for generic $\k$. The other symmetries are shared between the LO and HO models. 
\\ \\
The parameters for Eq.~\eqref{eq:tl}-\eqref{eq:dt} are documented in Tables XV-XVII of Ref.~\cite{zhang2024universal} and on Github \cite{zhang_Universal_2024github}.
The band structure comparison between the ``full'' ($77$-parameter) model in Ref.~\cite{zhang2024universal} and our reduced ($16$-parameter) model is shown in Fig.~\ref{fig:365} for $\theta=3.65^{\circ}$. We see that the reduced model reproduces the bands of the full model with very high accuracy.

\subsection{Valley Chern numbers}
\begin{figure}
    \centering
    \includegraphics[width=\linewidth]{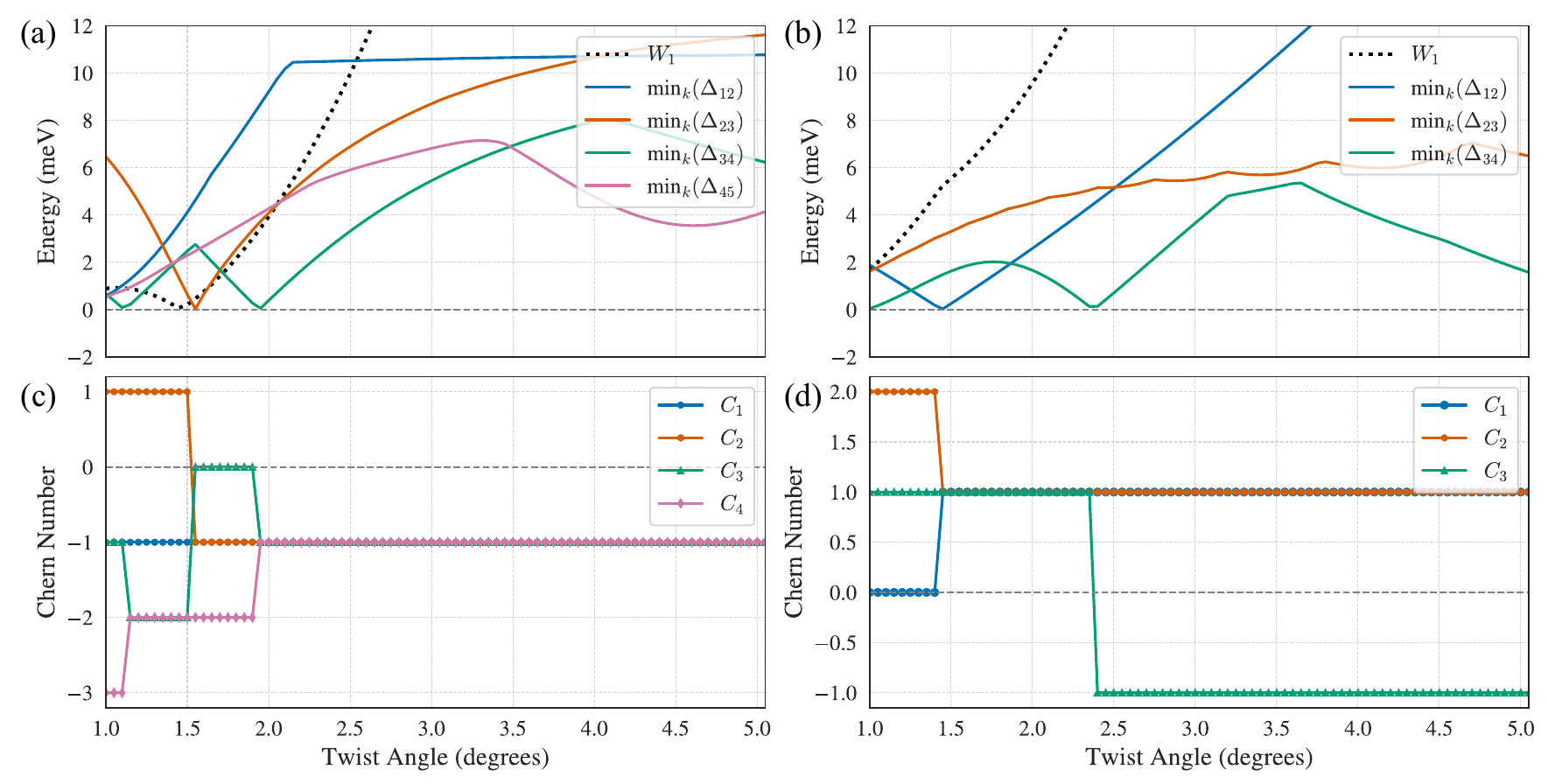}
    \caption{(a - b) Bandwidth of the topmost band $W_1$ and minimum energy distances between neighboring bands $\Delta_{ij}(\k) = E_i(\k)- E_j(\k)$ for the LO and HO model respectively.  (c - d) Chern number $C_i$ of the topmost $i$-th band for the LO and HO model respectively.}
    \label{fig:gapChern}
\end{figure}

To characterize the topological properties of the wavefunctions in the continuum models, we compute the Chern numbers of the topmost, relatively isolated bands. The degree of isolation is determined by the minimum energy separation between adjacent bands, $\Delta_{ij}(\k) = E_i(\k)- E_j(\k)$. As shown in Fig.~\ref{fig:gapChern}, the Chern numbers obtained from our reduced model, even when applied outside its formal range of validity, are consistent with those reported in Ref.~\cite{zhang2024universal} (with the exception of the case at $\theta=2.13^{\circ}$).

We find that for both models, the total Chern number of these topmost bands is non-zero. This net topological charge obstructs the construction of a simple tight-binding model based on exponentially localized Wannier functions. We also note that the emergence of flat bands (i.e., the ``magic'' condition where bandwidth $W_1 = 0$) is a model-dependent feature.

\section{Interacting model}
The continuum Hamiltonian can be rewritten in momentum space as
\begin{eqnarray}
H_0 & = & \sum_{\k\in \text{mBZ}} \sum_{\G,\G'} \sum_{m,n=t,b} \psi^\dagger_{m,\k+\G}\left[H(\k)\right]_{m\G,n\G'} \psi_{n,\k+\G'} \\
 & =: & \sum_{\k\in \text{mBZ}} \psi^\dagger_\k H(\k) \psi_\k \,,
\end{eqnarray}
where mBZ stands for the moir\'e or mini Brillouin zone. The eigenvectors of $H(\k)$ will be denoted as $|u_\alpha(\k)\rangle$. 
\\ \\
The screened Coulomb interaction is given by
\begin{eqnarray}
H_C & = & \frac{1}{2}\int\mathrm{d}^2\r\int\mathrm{d}^2\r'\, V(\r-\r')\psi^\dagger_m(\r)\psi^\dagger_n(\r')\psi_n(\r')\psi_m(\r) \\
 & = & \frac{1}{2A}\sum_{\k,\k',\q}V(\q)\psi^\dagger_{m,\k-\q}\psi^\dagger_{n,\k'+\q}\psi_{n,\k'}\psi_{m,\k}
\end{eqnarray}
Here summation over repeated indices is implicit, $A$ is the sample area, and
\begin{equation}
V(\q) = \int\mathrm{d}^2\r\,V(\r)e^{i\q\cdot\r} = \frac{e^2}{2\epsilon_0\epsilon_r}\frac{\tanh(qD)}{q}\,,
\end{equation}
with $D$ the screening length induced by the gates. We now rewrite the sums over $\k$ and $\k'$ as
\begin{equation}
\sum_\k \rightarrow \sum_{\k\in \text{mBZ}}\sum_{\tau=\pm}\sum_\G\,,
\end{equation}
with $\tau$ labeling the two different valleys at the K-points, and we approximate the Coulomb interaction as
\begin{equation}
H_C = \frac{1}{2A}\sum_\q \sum_{\k,\k'\in\text{mBZ}}\sum_{\tau,\tau'}\sum_{\G,\G'}V(\q)\psi^\dagger_{m,\tau,\G}(\k-\q)\psi^\dagger_{n,\tau',\G'}(\k'+\q)\psi_{n,\tau',\G'}(\k')\psi_{m,\tau,\G}(\k)
\end{equation}
Here we have defined $\psi^\dagger_{n,\tau,\G}(\k) = \psi^\dagger_n(\k+\tau\K_{\tau}+\G)$ with $\K_{\tau}$ the approximate location of the mBZ Gamma point relative to the mono-layer WSe$_2$ Gamma point. The above expression is approximate because we have neglected inter-valley scattering, which is suppressed due to the long-range nature of the interaction by a factor of $\sim V(2\K_{\tau})/V(0)$.
\\ \\
Going to the band basis the Coulomb interaction becomes
\begin{equation}
H_C = \frac{1}{2A}\sum_\q \sum_{\k,\k'\in\text{mBZ}} \sum_{\tau,\tau'}  V(\q) \left[ \Lambda^\tau_\q(\k)\right]_{\alpha\beta} \left[ \Lambda^{\tau'}_{-\q}(\k')\right]_{\lambda\sigma} c^\dagger_{\alpha,\k-\q} c^\dagger_{\lambda,\k'+\q} c_{\sigma,\k'}c_{\beta,\k}
\end{equation}
The form factors are defined as
\begin{equation}
\left[\Lambda^\tau_\q(\k)\right]_{\alpha\beta} = \langle u_{\tau,\alpha}(\k-\q)|u_{\tau,\beta}(\k)\rangle\,,
\end{equation}
where $|u_{\tau,\alpha}(\k)\rangle = |u_\alpha(\k)\rangle$ if $\tau = +$, and $|u_{\tau,\alpha}(\k)\rangle = |u_\alpha(-\k)\rangle^*$ if $\tau = -$ as dictated by time-reversal, which interchanges the two valleys.
\\ \\
An important subtlety is that the original continuum model is obtained by fitting to DFT band structures, which already incorporate Coulomb interaction effects. So if we add back the Coulomb interaction to the continuum model, then we are double-counting Coulomb contributions. To remedy this, we take the complete Hamiltonian to be
\begin{equation}
H = H_0 - H_{\text{sub}} + H_C\,,
\end{equation}
where $H_{\text{sub}}$ is a quadratic subtraction Hamiltonian which ensures that the continuum model bands are an exact solution of the Hartree-Fock self-consistency equations. So we have to take $H_{\text{sub}} = H_h[P_0] + H_f[P_0]$, where $\left[P_0(\k)\right]_{\alpha\beta} = \delta_{\alpha\beta}$ is the density matrix of the fully filled continuum bands, and
\begin{eqnarray}
H_h[P] & = & \frac{1}{A}\sum_\G V(\G)\left[\sum_{\k'}\text{tr}\left(P(\k)\Lambda_\G(\k') \right) \right]\sum_\k c^\dagger_\k \Lambda_{-\G}(\k)c_\k \\
H_f[P] & = & -\frac{1}{A}\sum_\k  c^\dagger_\k\left[\sum_\q V(\q) \Lambda_{-\q}(\k-\q)P(\k-\q)\Lambda_\q(\k)\right]c_\k
\end{eqnarray}
are the Hartree and Fock potentials.
\\ \\
As a final step we band-project. This means that choose a small set of active bands in each valley, which are closest to the Fermi energy. All the other remote bands are `frozen out', meaning that we take these bands to be completely filled and hence we do not allow them to hybridize with the active bands. The Coulomb interaction contains terms comprising entirely of electron operators in the active bands, terms with electron operators entirely in the remote bands, and terms with both electron operators in the active bands and in the remote bands. When we project the Coulomb interaction in the subspace with frozen remote bands, the latter type of mixed terms (i.e. containing electrons in both active and remote bands) will generate Hartree and Fock potentials for the electrons in the active bands. Specifically, if we define 
\begin{equation}
\left[P_r(\k) \right]_{\alpha\beta} = \begin{cases}  \delta_{\alpha\beta} & \text{if $\alpha,\beta$ \text{are remote bands}} \\  0 & \text{otherwise} \end{cases}
\end{equation}
as the density matrix of the frozen remote bands, then projecting the Coulomb interaction produces the quadratic terms $H_h[P_r]+H_f[P_r]$ for the electrons in the active bands. The projected Hamiltonian is thus given by
\begin{equation}
H_P = H_0 + H_h[P_r] + H_f[P_r] - H_{\text{sub}}  + H_C\,,
\end{equation}
where now all electron operators live exclusively in the active bands. We can further simplify this expression by noting that
\begin{equation}
H_h[P_r] + H_f[P_r] - H_{\text{sub}} = H_h[P_r-P_0] + H_f[P_r-P_0] = -H_h[P_a] - H_f[P_a]\,,
\end{equation}
where $P_a = P_0 - P_r$ is the density matrix of the completely filled active bands. So the final form of the Hamiltonian for the electrons in the active bands is
\begin{equation}
H_P = H_0 - H'_{\text{sub}} + H_C\,,
\end{equation}
with $H'_{\text{sub}} = H_h[P_a] + H_f[P_a]$.

\section{Real space IVC order parameter}
The real-space IVC order parameter in top and bottom layers can expressed as
\begin{equation}
    \langle \psi^{\dagger}_{l,\tau=+}(\r)\psi_{l,\tau=-}(\r)\rangle = |M^{l}_{xy}(\r)|e^{i\tilde{\theta}_{l}(\r)},
\end{equation}
with $l=t,b$ and
\begin{subequations}
\begin{align}
    & \psi^{\dagger}_{l,\tau=+}(\r) = \sum_{\k\in\text{mBZ}}\sum_{\G}\sum_{\alpha}e^{-i(\K_{l} + \k+\G-\frac{\boldsymbol{\kappa}_+}{2})\cdot\r} u^{\tau=+}_{\k,\G,l,\alpha}c^{\dagger}_{\tau=+,\k,\alpha} = \sum_{\k\in\text{mBZ}}\sum_{\alpha}e^{-i(\K_{l}+\k-\frac{\boldsymbol{\kappa}_+}{2})\cdot\r} \varphi^{\tau=+}_{\k,l,\alpha}(\r)c^{\dagger}_{\tau=+,\k,\alpha}\\
    & \psi_{l,\tau=-}(\r) = \sum_{\k\in\text{mBZ}}\sum_{\G}\sum_{\alpha}e^{i(-\K_{l} + \k+\G+\frac{\boldsymbol{\kappa}_+}{2})\cdot\r} \left[ u^{\tau=-}_{\k,\G,l,\alpha} \right]^{*}c_{\tau=-,\k,\alpha} = \sum_{\k\in\text{mBZ}}\sum_{\alpha}e^{i(-\K_{l}+\k+\frac{\boldsymbol{\kappa}_+}{2})\cdot\mathbf{r}} \left[ \varphi^{\tau=-}_{\k,l,\alpha}(\r) \right]^{*}c_{\tau=-,\k,\alpha} .
\end{align}
\end{subequations}
Note that in this appendix, we lighten the notation by replacing $\tilde{\k}$, which labels eigenvalues of the generalized translation symmetry $T'_{\a_i}$ as defined in the main text, with $\k$. This explains the presence of the shifts by $\pm \boldsymbol{\kappa}_+/2$ in the plane wave factors. The wavefunctions in $+$ and $-$ valley are related by time-reversal symmetry as
\begin{equation}
    \varphi^{\tau=-}_{\k,l,\alpha}(\r) = \varphi^{\tau=+^{*}}_{-\k,l,\alpha}(\r) = \mathcal{T}'\left[\varphi^{\tau=+}_{\k,l,\alpha}(\r)\right]
\end{equation}
For the $T_{\a_i}'$ translation symmetric state, the expression for the real space order parameter becomes
\begin{equation}
\begin{split}
    |M^{l}_{xy}(\r)|e^{i\tilde{\theta}_{l}(\r)} &= e^{-2i\K_{l}\cdot\r}e^{i\boldsymbol{\kappa}_+\cdot\r}\sum_{\k,\k'\in\text{mBZ}}\sum_{\alpha,\beta}e^{-i(\k-\k')\cdot\r}\left[\varphi^{\tau=-}_{\k',l,\alpha}(\r) \right]^{*} \langle c^{\dagger}_{\tau=+,\k,\beta}c_{\tau=-,\k',\alpha}\rangle\varphi^{\tau=+}_{\k,l,\beta}(\r) \\
    &= e^{-2i\K_{l}\cdot\r}e^{i\boldsymbol{\kappa}_+\cdot\r}\sum_{\k\in\text{mBZ}}\sum_{\alpha,\beta}\varphi^{\tau=-^{*}}_{\k,l,\alpha}(\r)\rho^{-+}_{\alpha,\beta}(\k)\varphi^{\tau=+}_{\k,l,\beta}(\r).
\end{split}
\end{equation}
As final step we remove the rapidly oscillating contribution due to the factor $e^{2i\K_{l}.\r}$ and absorb it into the phase $\tilde{\theta}_l(\mathbf{r})$. The resulting ‘slow’ oscillating phase $\propto e^{i\theta_l(\mathbf{r})}$ then characterizes the IVC texture at the moiré scale. The final expression becomes
\begin{equation}
    |M^{l}_{xy}(\r)|e^{i\theta_{l}(\r)} = e^{i\boldsymbol{\kappa}_+\cdot\r}\sum_{\k\in\text{mBZ}}\sum_{\alpha,\beta}\varphi^{\tau=-^{*}}_{\k,l,\alpha}(\r)P^{-+}_{\alpha,\beta}(\k)\varphi^{\tau=+}_{\k,l,\beta}(\r),
\end{equation}
with
\begin{equation}
    e^{i\theta_{l}(\r)} = e^{i\tilde{\theta}_{l}(\r)}e^{2i\K_{l}\cdot\r}.
\end{equation}
The angle $\theta_{l}(\r)$ is what we call the (in-plane) IVC-angle in the main text.

When the $T'_{\a_i}$ translation symmetry is broken in the multi-Q states, the Brillouin zone folds along $\frac{\G_{1}}{2}$ and $\frac{\G_{3}}{2}$, requiring an additional set of band indices $n,m \in {0,1}$. The expression for the real-space order parameter in this case becomes
\begin{equation}
\begin{split}
    \hspace{-5mm} |M^{l}_{xy}(\r)|e^{i\theta_{l}(\r)} &= e^{i\boldsymbol{\kappa}_+\cdot\r}\sum_{\k,\k'\in \text{fmBZ}}\sum_{\alpha,\beta}\sum_{\substack{n,n' \\ m,m'}}\bigg(
e^{-i(\k-\k'+(n-n')\frac{\G_{1}}{2}+(m-m')\frac{\G_{3}}{2})\cdot\r} \times\\
&  \hspace{4 cm}\left[\varphi^{\tau=-}_{\k',n',m',l,\alpha}(\r) \right]^{*} \langle c^{\dagger}_{\tau=+,\k,n,m,\beta}c_{\tau=-,\k',n',m',\alpha}\rangle\varphi^{\tau=+}_{\k,n,m,l,\beta}(\r) \bigg) \\
    &= e^{i\boldsymbol{\kappa}_+\cdot\r}\sum_{\k\in\text{fmBZ}}\sum_{\alpha,\beta}\sum_{\substack{n,n' \\ m,m'}}
e^{i((n'-n)\frac{\G_{1}}{2}+(m'-m)\frac{\G_{3}}{2})\cdot\r}\left[\varphi^{\tau=-}_{\k,n',m',l,\alpha}(\r)\right]^{*}P^{-+}_{\alpha,n',m';\beta,n,m}(\k)\varphi^{\tau=+}_{\k,n,m,l,\beta}(\r),
\end{split}
\end{equation}
where fmBZ stands for folded moiré Brillouin zone, and we already absorbed the fast oscillating phase factor as before.

\section{Results for LO model}\label{app:LOresults}
In this section we present the numerical results for the LO model and compare them with those of the HO model discussed in the main text. Although the LO parameters were extracted from DFT calculations for a $5^{\circ}$ twist-angle structure, we find that the two models exhibit remarkably similar features.
\begin{figure*}[h]
    \centering
    \includegraphics[width=\linewidth]{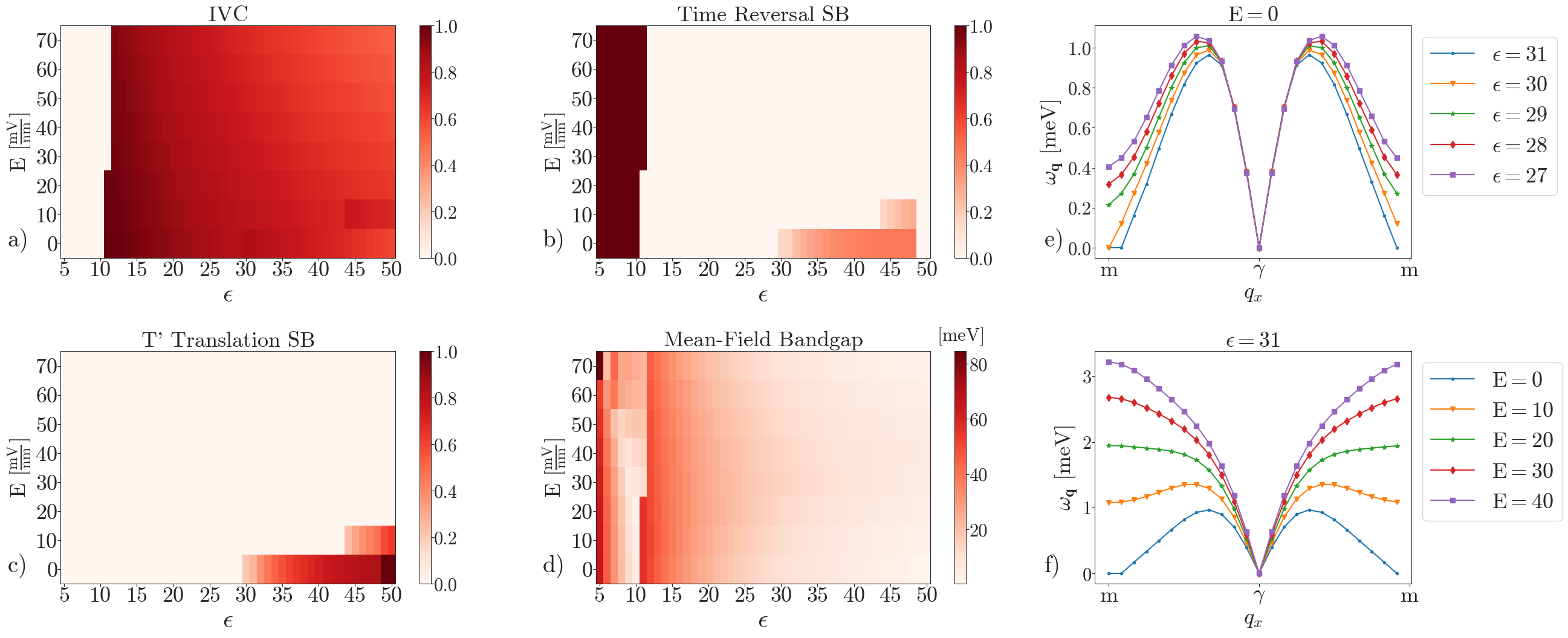}
    \caption{Numerical results for LO model obtained on a $24\times 24$ \moire lattice. (a) Spin-valley U(1) order parameter $O_{\text{IVC}}$, (b) $\mathcal{T}'$ time-reversal order parameter $O_{\mathcal{T}'}$, (c) order parameter $O_{T'}$ for the generalized translation symmetry $T'_{\a_i}$. (d) Mean-field bandgap. (e-f) TDHF Goldstone mode dispersion relation $\omega_\q$ along $\q=(q_x,q_y=0)$, both for fixed $E$ as a function of $\epsilon$ (e), and for fixed $\epsilon$ as a function of $E$ (f).}
    \label{fig:BigFig_LO}
\end{figure*}

The results in Fig.~\ref{fig:BigFig_LO} show that even for the simplified LO model we still find both the non-coplanar and coplanar multi-Q states. In the LO model, the $120^{\circ}$ AFM–to–non-coplanar multi-Q transition shifts to somewhat larger $\epsilon=30$ (compared to $\epsilon=24$ in the HO model). This can also be seen from the softening of the collective modes near the $M$-points. 

\begin{figure*}
    \centering
    \includegraphics[width=\linewidth]{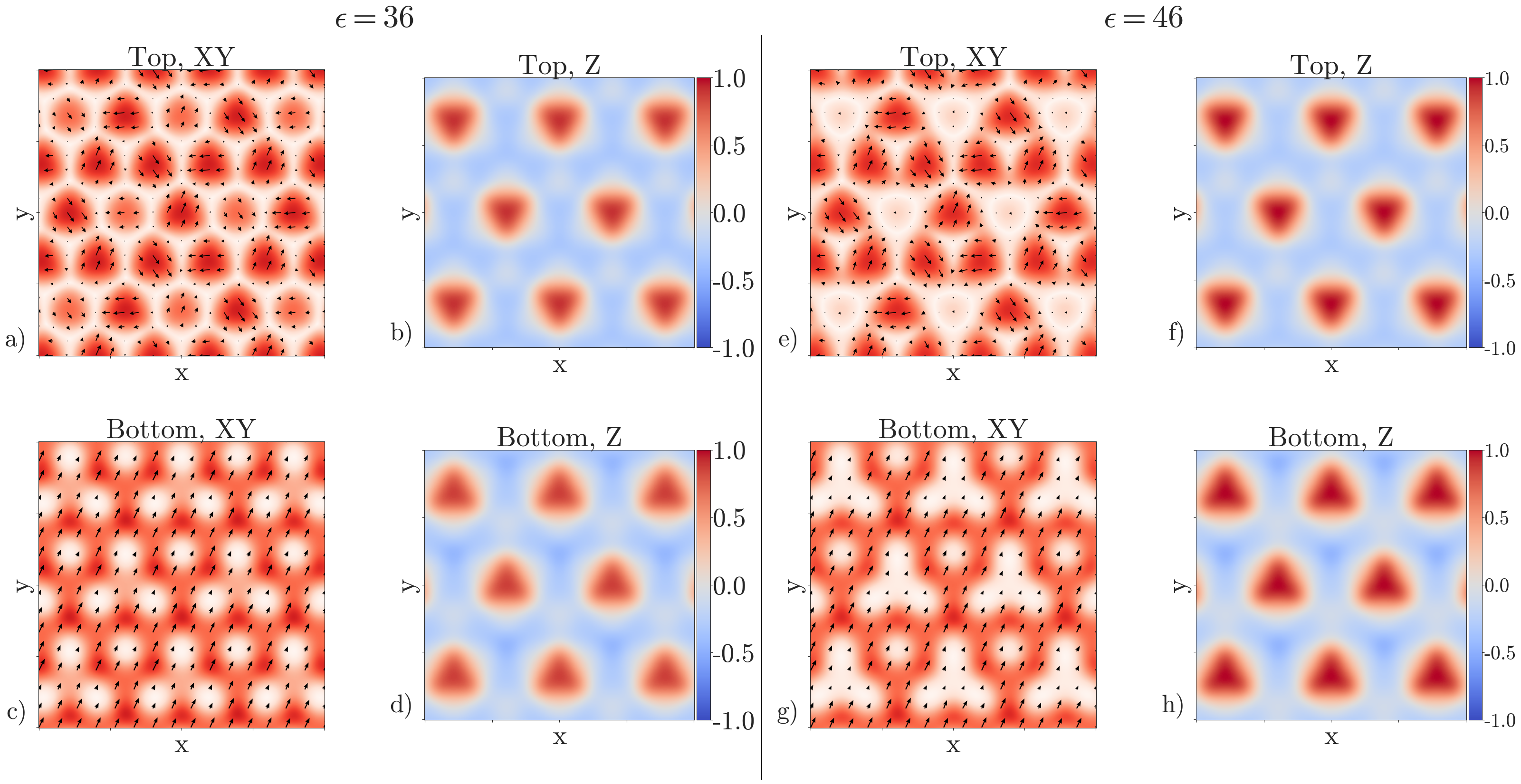}
    \caption{(a-d) Non-coplanar multi-Q spin texture in the top and bottom layers at $\epsilon = 36$. (a) In-plane magnetic order in the top layer, described by $|M_{xy}^t(\r)|$ (color) and $\theta_t(\r)$ (arrows). (b) $M_z^t(\r)$, the out-of-plane component of magnetic order in the top layer. (c) In-plane magnetic order in the bottom layer, described by $|M_{xy}^b(\r))|$ (color) and $\theta_b(\r)$ (arrows). (d) $M_z^b(\r)$, the out-of-plane component of magnetic order in bottom layer. (e-h) Same as (a-d), but for $\epsilon=46$.}
    \label{fig:spintext_LO}
\end{figure*}

In the LO model, the coplanar multi-Q phase is only found at very high dielectric constants, $\epsilon=49$–$50$. The key difference between the two models is that for the HO model the coplanar multi-Q phase is present in a larger region of the phase diagram. Spin textures for the non-coplanar state in the LO model are shown in Figure \ref{fig:spintext_LO} for two representative values of $\epsilon$.

Figure \ref{fig:spintext_LO} shows that near the transition at $\epsilon=30$, the in-plane spin texture in top layer again closely resembles that of the $120^{\circ}$ AFM. Deeper in the non-coplanar multi-Q phase, the texture deforms in a similar way as observed in the HO model. Throughout the entire non-coplanar region, the bottom layer exhibits a nearly constant in-plane spin direction $\theta_l(\r)$. In addition, a pronounced out-of-plane spin component is found with a period of two \moire lattice constants, which grows in magnitude deeper in the non-coplanar phase. Note that differences in the in-plane spin orientation for the LO and HO spin textures arise purely from spontaneous symmetry breaking and do not have physical significance.

\begin{figure}[h]
    \centering
    \includegraphics[width=0.75\linewidth]{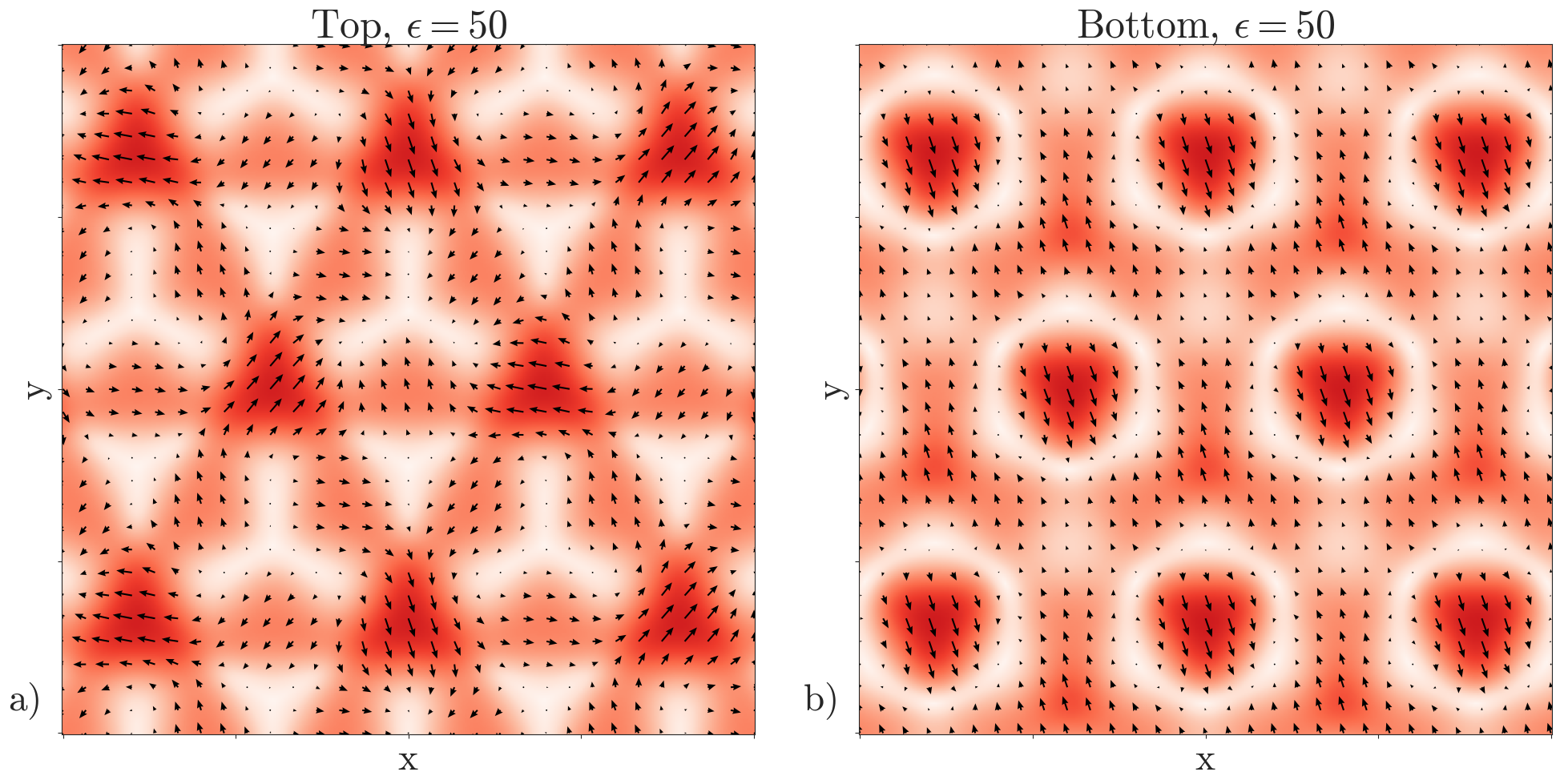}
    \caption{In-plane spin texture for the coplanar multi-Q state at $\epsilon=50$ in (a) top layer and (b) bottom layer. Color represents $|M_{xy}^l(\r)|$, and arrows represent $\theta_l(\r)$.}
    \label{fig:LOcoplanar_XY}
\end{figure}

In the coplanar multi-Q state shown in Figure \ref{fig:LOcoplanar_XY}, the top-layer spin texture again resembles that of the $120^{\circ}$AFM, but now formed on unit cells separated by two moiré lattice constants. Also for the LO model, the in-plane spin direction in the bottom layer is no longer constant in the coplanar multi-Q state. 

Taken together, the results for the LO and HO models demonstrate that the emergence of coplanar and non-coplanar multi-Q states is a robust feature of twisted WSe$_2$, i.e. it is insensitive to microscopic details of the continuum model.

\end{document}